\documentclass[11pt]{article}
\usepackage[pdftex]{graphicx}
\usepackage{hyperref}
\usepackage{cite}
\usepackage{bm}
\usepackage{amsmath}
\usepackage{amsmath,amsfonts,amssymb,amsthm,nccmath,latexsym,mathtools}
\usepackage{xcolor}
\usepackage{amsmath}
\usepackage{cancel} 
\usepackage{amssymb}
\usepackage{amsbsy} 
\usepackage{footnote}

\hypersetup{
	colorlinks   = true, 
	urlcolor     = blue, 
	linkcolor    = blue, 
	citecolor   = red 
}
 
\def\o{\omega}

\def\no{\nonumber}

\def\d{\delta}

\def\p{\partial}

\def\na{\nabla}
\def\T{\Theta}
\def\lie{\pounds_{\xi}}
\def\t{\tilde}

\def\L{\Lambda}
\def\ML{\mathcal{L}}
\def\ME{\mathcal{E}}
\def\be{\begin{equation}}
\def\ee{\end{equation}}
\def\ba{\begin{align}}
\def\ea{\end{align}}

\def\mg{\sqrt{-g}}

\def\U{\mathcal{U}}
\def\V{\mathcal{V}}
\def\J{\mathcal{J}}

\begin{document}
\title{{\bf{\Large Extended phase space thermodynamics of black holes: A study in Einstein's gravity and beyond}}}
\author{
{\bf{\normalsize Krishnakanta Bhattacharya}}\footnote {\color{blue} krish.phys@gmail.com}\\
IUCAA, Post Bag 4, Pune University Campus,
   Ganeshkhind, Pune 411 007,
   Maharashtra, India.}
\maketitle
\begin{abstract}
In the extended phase space approach, one can define thermodynamic pressure and volume that gives rise to the van der Waals type phase transition for black holes. For Einstein's GR, the expressions of these quantities are unanimously accepted. Of late, the van der Waals phase transition in black holes has been found in  modified theories of gravity as well, such as the $f(R)$ gravity and the scalar-tensor gravity. However, in the case of these modified theories of gravity, the expression of pressure (and, hence, volume) is not uniquely determined. In addition, for these modified theories, the extended phase space thermodynamics has not been studied extensively, especially in a covariant way. Since both the scalar-tensor and the $f(R)$ gravity can be discussed in the two conformally connected frames (the Jordan and the Einstein frame respectively), the arbitrariness in the expression of pressure, will act upon the equivalence of the thermodynamic parameters in the two frames. We highlight these issues in the paper. Before that, in Einstein's gravity (GR), we obtain a general expression of the equilibrium state version of first law and the Smarr-like formula from the Einstein's equation for a general static and spherically symmetric (SSS) metric. Unlike the existing formalisms in literature which defines thermodynamic potential in order to express the first law, here we directly obtain the first law as well as the Smarr-like formula in GR in terms of the parameters present in the metric (such as mass, charge \textit{etc.}). This study also shows how the extended phase space is formulated (by considering the cosmological constant as variable) and, also shows why the cosmological constant plays the role of thermodynamic pressure in GR in extended phase space. Moreover, obtaining the Smarr formula from the Einstein's equation for the SSS metric suggests that this dynamical equation encodes more information on BH thermodynamics than what has been anticipated before. 

\end{abstract}


\section{Introduction}
The connection between general relativity and thermodynamics has been found almost fifty years ago from the works of Bekenstein and Hawking \cite{Bekenstein:1973ur, Hawking:1974sw, Bardeen:1973gs}. Till date, several interesting results are coming up with the course of time and further consolidate the connection between the two. Black hole (BH) phase transition is one of the most interesting results in black hole mechanics which was obtained long ago by Davies \cite{Davies:1989ey}. It was then identified that the critical point of the phase transition as one where the heat capacity shows an infinite discontinuity. The presence of phase transition and the critical phenomena was later found, more prominently for the AdS black holes. Firstly it was found by Hawking and Page \cite{Hawking:1982dh}, which is known as the Hawking-Page phase transition. Later, the $P-V$ criticality of conventional thermodynamics was found in BH mechanics, where the thermodynamic pressure ($P$) was defined in two different ways. It was first noticed in the non-extended phase space approach \cite{Chamblin:1999tk, Chamblin:1999hg} where it was shown that the black hole undergoes a transition, which is similar to the van der Waals-Maxwell liquid-gas system, where the pressure was identified as the inverse of Hawking temperature. This approach was not accepted much due to the lack of proper canonical definition of pressure and volume. Later, in the extended phase space approach, the cosmological constant of the AdS black hole was treated as a variable and was found that the $P-V$ criticality exists in the black hole thermodynamic system of the AdS black holes and the pressure ($P$) was identified in terms of the cosmological constant \cite{Kastor:2009wy, Dolan:2010ha, Dolan:2011xt, Dolan:2011jm, Dolan:2012jh, Kubiznak:2012wp, Kubiznak:2016qmn}. This extended phase space formalism was widely accepted and $P-V$ criticality was later extensively studied with the identification of cosmological constant as the thermodynamic pressure (for general analysis in this topic, see \cite{Majhi:2016txt,Bhattacharya:2017hfj,Bhattacharya:2017nru,Bhattacharya:2019qxe,Dehyadegari:2018pkb,Bhattacharya:2020jgk}).

The extended phase space formalism, as discussed above, was mostly studied in Einstein's gravity (GR). Although, till date, GR is considered as the most viable theory to describe gravity, one cannot disregard the alternative theories of gravity for several obvious reasons. In addition, thermodynamic features of general relativity has been found in alternative gravity theories as well \cite{Eling:2006aw,Elizalde:2008pv,Chirco:2009dc,Padmanabhan:2002sha,Padmanabhan:2003gd,Paranjape:2006ca,Kothawala:2007em,Faraoni:2010yi,Bamba:2009gq}, which suggests that the thermodynamic property of gravity ranges well-beyond Einstein's GR (also see \cite{Addazi:2016hip, Bhattacharya:2016lup, Chakraborty:2016ydo, Sami:2017npa}, which discusses thermodynamic aspects of $f(R)$ theory and other modified theories of gravity). Interestingly, the $P-V$ criticality has been found in scalar-tensor (ST) / Brans-Dicke and $f(R)$ gravity as well \cite{Chen:2013ce,Ovgun:2017bgx,Hendi:2015kza,Hendi:2015hgg}. However, there seems to have a disparity in the expression of the thermodynamic pressure. It is well known, in general, $f(R)$ gravity can be studied equivalently as the scalar-tensor gravity with the identification $\phi=f'(R)$, where $\phi$ is the scalar field which is non-minimally coupled with the Ricci-scalar in the ST gravity and $f'(R)=\p f(R)/\p R$ in the $f(R)$ gravity. Now, the expression of pressure has been considered as different in different papers. For example, in \cite{Chen:2013ce}, the thermodynamic pressure is identified in terms of the cosmological constant multiplied with the factor $f'(R)$ (or the conformal factor) and, on the other hand, in \cite{Ovgun:2017bgx, Hendi:2015kza} the cosmological constant itself is identified as the thermodynamic pressure in $f(R)$ gravity and in Brans-Dicke (BD) gravity. Apart from this difference in the expression of thermodynamic pressure, there is another important aspect.  As we have mentioned, both $f(R)$ and BD gravity can be studied as a special case of scalar-tensor gravity and can be described in two conformally connected frames \textit{i.e.}, Jordan and Einsten frame. In earlier works the equivalence of thermodynamic parameters, in the two conformally connected frames, has been established by us and other groups \cite{Koga:1998un,Bhattacharya:2017pqc,Bhattacharya:2018xlq,Dey:2021rke,Bhattacharya:2020jgk}. But, those analysis were done in the non-extended phase space approach as the variation of the cosmological constant was not accounted. Therefore, the question arises how the thermodynamic parameters, defined in the extended phase space approach, are related in the two conformally connected frames. Furthermore, the question arises: whether this arbitrariness in the definition of pressure influences the equivalence of the thermodynamic parameters in the two frames. 

 In this work, we address these issues. The paper will be described in two parts. In the first part (section \ref{SECSSS}), the entire discussion is on Einstein's GR, where the thermodynamics has been obtained only from the Einstein's equation for the SSS BH. In this case, we derive a novel approach and obtain the thermodynamic first law and the Smarr-like formula \cite{Smarr:1972kt} from the Einstein's equation for the SSS BH. The idea is, if a static and spherically symmetric solution of Einstein's equation depends upon several parameters which characterizes the solution space of the BH (such as mass, charge, AdS radius \textit{etc.}), using the presented approach, one can obtain the equilibrium state version of the first law and the Smarr-like formula in a general way in any arbitrary dimension for a SSS black hole. The entire formulation is general, as it can be applied for any static and spherically symmetric metric (with $g_{00}=-1/g_{rr}$). The fact, that the Smarr formula can be obtained from the Einstein's equation itself, implies that the equation encodes more information on BH thermodynamics than anticipated before as it has not been claimed earlier that the Einstein's equation manifests itself as the Smarr formula (to the best of my knowledge). This is important in studying gravity as an ``emergent phenomena''. Taking a particular example (RNAdS black hole), we show that if the cosmological constant of an AdS black hole is considered as a variable, it looks like thermodynamic pressure and its conjugate quantity appears to be the volume of the sphere, the radius of which is the horizon radius.  Thus, we show the formulation of extended phase space in GR.
 
  In the second part of the paper (discussed in the section \ref{SECECON} and \ref{SECST}), we have looked beyond the Einstein's gravity. We have mentioned that the $P-V$ criticality has been found in literature for $f(R)$ and Brans-Dicke theory, both of which can be studied as the special case of the ST gravity. However, in literature, one can find that there is a disparity  regarding the expression of thermodynamic pressure and volume in literature for those cases. In addition, there are debates regarding the physical equivalence of the two conformally connected frames \cite{Faraoni:1999hp, ALAL,Faraoni:1998qx, Faraoni:2010yi, Faraoni:2006fx, Saltas:2010ga, Capozziello:2010sc, Padilla:2012ze, Koga:1998un, Jacobson:1993pf, Kang:1996rj, Deser:2006gt, Dehghani:2006xt, Sheykhi:2009vc,Steinwachs:2011zs, Kamenshchik:2014waa, Banerjee:2016lco, Pandey:2016unk, Ruf:2017xon,Karam:2017zno,Bahamonde:2017kbs,Karam:2018squ,Bhattacharya:2017pqc,Bhattacharya:2018xlq,Bhattacharya:2020wdl,Bhattacharya:2020jgk,Dey:2021rke,Capozziello:2006dj,Elizalde:2020icc,Bahamonde:2016wmz}. However, it has been found that the thermodynamic parameters (of non-extended phase space) are equivalent in the two conformally connected frames \cite{Koga:1998un,Bhattacharya:2017pqc,Bhattacharya:2018xlq,Dey:2021rke,Bhattacharya:2020jgk}. Therefore, in the extended phase space approach, here we study the connection of thermodynamic parameters in the two frames (especially pressure and volume). To address this issue in a covariant manner, we shall adopt the covariant Noether current formalism developed by Iyer and Wald\cite{Wald:1993nt,Iyer:1994ys}. Firstly, we introduce the conformal frame of GR, which corresponds to a particular case of Jordan frame of ST gravity. Then, following Iyer-Wald formalism, we define the thermodynamic quantities and obtain the first law considering cosmological constant as a variable. We then compare the thermodynamic parameters of the conformal frame (or the Jordan frame) with the thermodynamic parameters of GR (corresponding to the Einstein frame). We do this because, the thermodynamic parameters in GR is well defined. Therefore, it is simpler to compare the result with its conformal frame. After this, we generalize this result for the general scalar-tensor gravity in its two frames. Our analysis, not only proves the first law in the two frames, it provides the expressions of all thermodynamic parameters in a covariant way, especially thermodynamic volume which has not been shown for alternative theories of gravity, especially ones with the non-minimal coupling. Also, we discuss that the arbitrariness in the expression of pressure indeed influences the equivalence of the thermodynamic parameters.
 
The discussion of the paper is presented as follows. In the following section, we discuss our formalism to obtain first law and the Smarr-like formula in a general way for GR from the Einstein's equation. Here we also show how the identification of pressure in terms of cosmological constant gets justified. In section \ref{SECECON}, we adopt Wald's formalism and compare the thermodynamic parameters of Einstein's gravity with its conformal frame. Finally, we generalize these results for ST gravity in section \ref{SECST}. We provide conclusion and outlook of our analysis in section \ref{SECCON}.

\textit{Units and notations:} In this article, we have adopted geometrized units and have set $c$, $\hbar$ and $G$ as unity. Quantities with a bar overhead (such as $\bar A$) correspond to the of Einstein's GR. Quantities with a tilde overhead (such as $\t A$) correspond to the Einstein frame of ST gravity. For Jordan frame, or for the conformal frame of Einstein’s GR, we use plain notations (such as $A$).

\section{Thermodynamic first law and Smarr-like formula from the spacetime metric in GR: A general analysis in spherically symmetric spacetime} \label{SECSSS}
It is well-known that Einstein’s equation manifests itself as a thermodynamic law when it is projected
on the black hole horizon \cite{Hayward:1997jp, Padmanabhan:2002sha,Padmanabhan:2003gd,Paranjape:2006ca,Kothawala:2007em,Bhattacharya:2016kbm,Hansen:2016gud}. Here we have briefly mentioned two existing formalisms in literature to obtain
thermodynamic first law from Einstein’s equation for static and spherically symmetric spacetime.
Then we discuss our general formalism of obtaining first law and the Smarr-like formula from the Einstein’s
equation. We show that, in the existing formalisms, the first law is defined in terms of thermodynamic potential. In our case, the first law appears as the change of the BH entropy in terms of the “virtual”
change of the metric parameters (mass, charge etc.).

A general static and spherically symmetric metric is given as
\begin{align}
d\bar s^2=-\bar f(\bar r)d\bar t^2+\frac{d\bar r^2}{\bar f(\bar r)}+\bar r^2(d\bar \theta^2+\sin^2\bar \theta d\bar \phi^2)~, \label{SSSMET}
\end{align}
with $\bar f(\bar r)$ being the general function of $\bar r$. This metric satisfies the radial Einstein's equation ($\bar G_{\bar r}^{\bar r}=8\pi \bar T_{\bar r}^{\bar r}$), which is provided as
\begin{align}
\frac{-1+\bar f(\bar r)}{\bar r^2}+\frac{f'(\bar r)}{\bar r}=8\pi \bar T_{\bar r}^{\bar r}~. \label{EINEQ}
\end{align}
There are two routes to obtain the first law from the above Einstein's equation as described in literature. On the first route \cite{Padmanabhan:2002sha,Padmanabhan:2003gd}, the Einstein's equation \eqref{EINEQ} is cast on the horizon $\bar r=\bar r_H$ (where $\bar r_H$ is obtained using $\bar f(\bar r_H)=0$). Then both sides of \eqref{EINEQ} are multiplied with $(\bar r_H^2d\bar r_H)/2$ to obtain
\begin{align}
\bar Td\bar S-d\bar E=\bar P_{(r)}d\bar V~.\label{1STPADDY}
\end{align}
 The above equation is identified as the first law of a black hole, the horizon of which has undergone a virtual displacement from $\bar r_H$ to $\bar r_H+d\bar r_H$. Here $\bar P_{(r)}$ is known as the radial pressure and $\bar V$ is the geometric volume of a sphere of radius $\bar r_H$. The thermodynamic quantities are identified as
\begin{align}
\bar P_{(r)}=\bar T_{\bar r}^{\bar r}\Big|_{\bar r_H}~, \ \ \ \bar V=\frac{4\pi}{3}\bar r_H^3~,\ \ \ \bar T=\frac{\bar f'(\bar r_H)}{4\pi}~, \ \ \ \bar S=\pi \bar r_H^2~, \ \ \ \bar E=\frac{\bar r_H}{2}~. \label{THERMPADDY}
\end{align}
There is another route \cite{Hansen:2016gud} to obtain thermodynamic law from the Einstein's equation \eqref{EINEQ} in the literature, which is as follows. The Eq. \eqref{EINEQ} can be written in terms of a generalized equation of state
\begin{align}
\bar P_{(r)}=B(\bar r_H)+C(\bar r_H)\bar T~,
\end{align} 
where the expression of $\bar P_{(r)}$ and $\bar T$ are given in \eqref{THERMPADDY}. Therefore, taking the variation on both sides and multiplying it with $\bar V$ leads to the thermodynamic relation
\begin{align}
\bar V\delta \bar P_{(r)}=\bar S\delta \bar T+\delta \bar G~,\label{1STMANN}
\end{align}
where, 
\begin{align}
\bar G=\int^{\bar r_H}d\bar x\bar V(\bar x)B'(\bar x)+\bar T\int^{\bar r_H}d\bar x\bar V(x)C'(\bar x)
\no 
\\
=\bar P_{(r)}\bar V-\bar T\bar S-\int^{\bar r_H}d\bar x\bar V'(\bar x)B(\bar x)~,
\no 
\\
\bar S=\int^{\bar r_H}d\bar x\bar V'(\bar x)C(\bar x)~.
\end{align}
In the present case (Einstein's gravity), $B(\bar r_H)=-(8\pi \bar r_H^2)^{-1}$ and $C(\bar r_H)=1/(2\bar r_H)$, implying 
\begin{align}
\bar G=\frac{\bar r_H}{3}(1-\pi \bar r_H \bar T)~, \label{GIBBS}
\end{align}
and $\bar S$ is given by quarter of the horizon area as given in \eqref{THERMPADDY}. Thus, the thermodynamic quantities are identified as the same in the two different routes. The identified internal energy ($\bar E$) of the first route and the Gibbs function ($\bar G$) as defined in the second route are connected to each other by the Legendre transformation $\bar E=\bar G+\bar T\bar S-\bar P_{(r)}\bar V$.

The second route \cite{Hansen:2016gud} is adopted (as it is mentioned there) in order to remove the ambiguity to distinguish the ``heat'' and ``work'' term of the first route \cite{Padmanabhan:2002sha,Padmanabhan:2003gd} as both $\bar S$ and $\bar V$ are functions of $\bar r_H$ only. However, in both the routes, the thermodynamic potentials (such as $\bar E$ defined in the first route and $\bar G$ in the second) are needed to be defined in order to match the derived relations with 1st law of conventional thermodynamics. Note that the defined thermodynamic potentials, as  defined in the two routes mentioned above, are not unique. For instance, for the Schwarzschild black hole, the internal energy ($\bar E$), which is defined as half of the horizon radius, coincides with the mass of the black hole ($\bar M$); while for Reissner N\"ordstrom (RN) black hole, $\bar E$ is not the same as $\bar M$. Furthermore, according to the no hair theorem, the parameters appearing in the black hole metric are more fundamental than the thermodynamic potentials, which are defined while obtaining the first law following these two routes. Moreover, as we have noted, the radial pressure is identified as $\bar P_{(r)}=\bar T^{\bar r}_{\bar r}(\bar r_H)$ in these methods. For a charged Reissner N\"ordstrom (RN) black hole in AdS spacetime, the radial pressure will be identified as $\bar P_{(r)}=\bar T^{\bar r}_{\bar r}=-\bar Q^2/(8\pi \bar r^4)-\bar \Lambda/8\pi$. However, for RN black hole in AdS space, the thermodynamic pressure is identified as $\bar P=-\bar \Lambda/8\pi$ which give rise to the $\bar P-\bar V$ criticality \cite{Kastor:2009wy, Dolan:2010ha, Dolan:2011xt, Dolan:2011jm, Dolan:2012jh, Kubiznak:2012wp, Kubiznak:2016qmn} and is consistent with the thermodynamics in the extended phase space. In other words, the radial pressure does not play the role of thermodynamic pressure for the $\bar P-\bar V$ criticality in the extended phase space.

Therefore, the natural question arises whether one can generally obtain the thermodynamic law (and the Smarr formula) of black holes in terms of the parameters ($\bar M$, $\bar Q$ \textit{etc.}) instead of defining any thermodynamic potential. More precisely, whether one can show the change of horizon area in terms of the change in parameters only. The formulation we developed in the following, requires Einstein's equation to be solved. However, the final form of the first law and the Smarr-like formula, requires only the spacetime metric and nothing else.

With the identification $\bar T^{\bar r}_{\bar r}=-\epsilon(\bar r)/(8\pi)$, the Einstein's equation \eqref{EINEQ} leads to the general solution \cite{paddybook}
\begin{eqnarray}
&&\bar f(\bar r)=1-\frac{\bar r_H}{\bar r}-\frac{1}{\bar r}\int_{\bar r_H}^{\bar r}\epsilon(x)x^2dx=1-\frac{1}{\bar r}\int^{\bar r}\epsilon(x)x^2dx-\frac{\bar r_H}{\bar r}\Big[1-\frac{1}{\bar r_H}\int^{\bar r_H}\epsilon(x)x^2dx\Big]~,
\end{eqnarray}
which implies
\begin{align}
\bar f(\bar r)=g(\bar r)-\frac{C}{\bar r}~,\label{FR}
\end{align}
where $C$ is the integration constant and
\begin{align}
g(y)=1-\frac{1}{y}\int^y\epsilon(x)x^2dx~. \label{GR}
\end{align}
The integration constant $C$ can be identified as the mass of the black holes for the following reasons. The Birkhoff's theorem \cite{BIRKHOFF} says that the exterior solution of a static and spherically symmetric metric without any source is given by the Schwarzschild metric. From  \eqref{GR}, we obtain $g(\bar r)=1$ and $\bar f(\bar r)=1-C/\bar r$ for $\epsilon(\bar r)=0$. Thus, we obtain the value of the integration constant as
\begin{align}
C=2\bar M=\bar r_Hg(\bar r_H)~,\label{MASS}
\end{align}
 and, thereby, we obtain
\begin{align}
\bar f(\bar r)=g(\bar r)-\frac{2\bar M}{\bar r}~. \label{FR2}
\end{align}

As we have mentioned earlier, for the spherically symmetric spacetime of the form \eqref{SSSMET}, the Einstein's equation is first order differential equation. Therefore, in the solution \eqref{FR2}, we have one integration constant, which is defined as the mass parameter of BH. However, in the spacetime metric, we can have other parameters (such as $\bar Q$) coming from the component of energy-momentum tensor $\epsilon$ (such as, for RN black hole, $\epsilon(\bar r)=\bar Q^2/\bar r^4$). Thus, we can write $g(\bar r)\equiv g(\bar r,\bar \mu^i)$, where $\bar \mu^i$ are the parameters (other than the mass) that appear in the spacetime metric and the number of parameters are kept arbitrary. In addition, we express $\bar f(\bar r)$ as $\bar f(\bar r,\bar M,\bar \mu^i)$.

\subsection*{Obtaining the first law}
Let us now consider a small virtual change of parameters, leading to a small (virtual) change in the horizon radius $\delta \bar r_H$\footnote{by virtual change, we mean that the change in parameters and horizon radius are not happening by any physical process. We just compare two BH solutions with radii $\bar r_H$ (mass $\bar M$ and so on) and $\bar r_H+\d \bar r_H$ (mass $\bar M+\d \bar M$ and so on).}. Therefore, from \eqref{MASS}, we obtain 
\begin{align}
2\delta \bar M=\Big[g(\bar r_H,\bar \mu^i)+\frac{\p g(\bar r_H,\bar \mu^i)}{\p \bar r_H}\bar r_H\Big]\d \bar r_H+\bar r_H\frac{\p g(\bar r_H,\bar \mu^i)}{\p\bar \mu^i}\d\bar \mu^i \label{delm}
\end{align}
Furthermore, from \eqref{FR} we obtain
\begin{align}
\bar f'(\bar r_H)=\frac{\p \bar f(\bar r)}{\p \bar r}\Big|_{\bar r=\bar r_H}=\frac{\p g(\bar r_H,\bar \mu^i)}{\p \bar r_H}+\frac{g(\bar r_H,\bar \mu^i)}{\bar r_H}~,\label{fprime}
\end{align}
In addition, from \eqref{FR2}, we obtain 
\begin{align}
\frac{\p g(\bar r_H,\bar \mu^i)}{\p\bar \mu^i}=\frac{\p \bar f(\bar r,\bar M,\bar \mu^i)}{\p\bar \mu^i}\Big|_{\bar r=\bar r_H}~.\label{gmu}
\end{align}
Substituting \eqref{fprime} and \eqref{gmu} in \eqref{delm}, we finally obtain
\begin{align}
\bar T\d \bar S=\d \bar M-\frac{\bar r_H}{2}\frac{\p \bar f(\bar r,\bar M,\bar \mu^i)}{\p\bar \mu^i}\Big|_{\bar r=\bar r_H}\d \bar \mu^i \label{1STKK}
\end{align}
This is the explicit from the first law which has been generally obtained. As we have promised, here the change in entropy of BH is obtained in terms of the change in parameters only. No thermodynamic potentials are needed to be defined. The identification of temperature ($\bar T=\bar f'(\bar r_H)/4\pi$) has been done in \eqref{1STKK} via standard arguments of thermal quantum field theory and we have used the well-known result of BH entropy being quarter of horizon area. More importantly, we need not know about the source explicitly (unlike the two routes mentioned earlier) to obtain the first law. If the spacetime metric is provided, we can obtain the thermodynamic law from \eqref{1STKK}.
\subsection*{Generalized Smarr-like formula}
Earlier, we have identified the mass of the black hole in terms of Eq. \eqref{MASS}. We replace $g(\bar r_H,\bar \mu^i)$ there using Eq. \eqref{fprime} and obtain
\begin{align}
\bar M=2\bar T\bar S-\frac{\bar r_H^2}{2}\frac{\p g(\bar r_H,\bar \mu^i)}{\p \bar r_H}~, \label{smarr}
\end{align}
where the expressions of $\bar T$ and $\bar S$ are mentioned earlier. The last term in Eq. \eqref{smarr} contributes when the source term (or the cosmological constant) is present. When there is no source (it corresponds to the Schwartschild case), $g(\bar r_H,\bar \mu^i)=1$ and we obtain $\bar M=2\bar T\bar S$, which is the Smarr formula for Schwartzschild black hole.

From Eq. \eqref{FR2}, we can define
\begin{align}
g(\bar r,\bar \mu^i)=\bar f(\bar r,\bar M,\bar \mu^i)\Big|_{\bar M=0}~. \label{gfconnection}
\end{align}
Therefore, using the above relation \eqref{gfconnection} in \eqref{smarr}, we obtain the Smarr-like formula as
\begin{align}
\bar M=2\bar T\bar S-\frac{\bar r_H^2}{2}\Big(\frac{\p \bar f(\bar r,\bar M,\bar \mu^i)}{\p \bar r}\Big)_{\bar r=\bar r_H,\bar M=0}~.\label{smarrf}
\end{align}

Thus, we show that for a spherically symmetric spacetime, if the metric is known, we can obtain the Smarr formula directly using Eq. \eqref{smarrf}.  Therefore, not only the first law, the Smarr formula can also be obtained from the Einstein's equation. 

\subsection*{Identification of thermodynamic pressure in extended phase space}
We now take a particular example to check whether the formulation, defined above, actually provides the first law for a known case. Let us consider charged spherically symmetric black hole solution with the cosmological constant. In that case, we know the metric solution (\textit{i.e.} the RN black hole in AdS space), which is given as
\begin{align}
\bar f(\bar r,\bar M,\bar \mu^i)=1-\frac{2\bar M}{\bar r}+\frac{\bar Q^2}{\bar r^2}-\frac{\bar \Lambda \bar r^2}{3}
\end{align}
Let us consider, that the horizon radius changes infinitesimally due to the change in $\bar M$, $\bar Q$ and $\bar \Lambda$ (\textit{i.e.} we consider the cosmological constant as a variable instead of being a true constant in the theory). Therefore, the first law, as provided by \eqref{1STKK}, is obtained as
\begin{align}
\d \bar M=\bar T\d \bar S+\frac{\bar Q}{\bar r_H}\d \bar Q-\frac{4\pi \bar r_H^3}{3}\d\Big(\frac{\bar \Lambda}{8\pi}\Big)~.\label{1STEXT}
\end{align}
One can identify $\bar Q/\bar r_H$ as the thermodynamic potential at the horizon ($\bar \Phi_H$). $4\pi \bar r_H^3/3$ corresponds to the volume of a sphere with radius $\bar r_H$. Therefore, in AdS spacetime (spacetime with the value of cosmological constant being negative), if the cosmological constant is considered as a variable, one can identify the last term of \eqref{1STEXT} as $\bar V\delta \bar P$ and the first law looks like
\begin{align}
\d \bar M=\bar T\d \bar S+\bar \Phi_H\d \bar Q+\bar V \d\bar P~, \label{1STEXT1}
\end{align}
where we have identified $\bar P=-\bar \Lambda/8\pi$. This identification of pressure (for GR) has been proved to be consistent in the literature. With these identifications the $\bar P-\bar V$ criticality has been obtained in BH thermodynamics \cite{Kubiznak:2012wp, Kubiznak:2016qmn}. It is noteworthy that, the BH mass ($\bar M$) can be identified as the enthalpy of the BH thermodynamic system \cite{Kastor:2009wy, Dolan:2010ha, Dolan:2011xt, Dolan:2011jm, Dolan:2012jh} when $\bar\L$ is considered as variable in the extended phase space approach. 

One can obtain the Smarr formula using our formulation. From Eq. \eqref{smarrf}, we obtain the Smarr formula of RN-AdS black hole as
\begin{align}
\bar M=2\bar T\bar S+\bar Q\bar \Phi_H-2\bar P\bar V~,
\end{align}
which agrees with the earlier results obtained in the literature \cite{Kastor:2009wy, Kubiznak:2012wp, Kubiznak:2016qmn}. However, the novelty in the present analysis is that we can obtain the Smarr formula directly from the metric.

\subsection*{Higher dimensional generalization of the first law and Smarr-like formula}
The formulation mentioned above, can be generalized to higher dimension as well. A general spherically symmetric metric in arbitrary $D$ spacetime dimension reads
\begin{align}
d\bar s^2=-\bar f(\bar r)d\bar t^2+\frac{d\bar r^2}{\bar f(\bar r)}+\bar r^2d\bar \Omega_{D-2}^2~. \label{metricd}
\end{align}
For this metric, we obtain the first law and the Smarr-like formula in a general way (see the appendix \ref{appen}). The first law is given as
\begin{align}
\d \bar M=\bar T\d \bar S+\frac{\bar r_H^{D-3}}{\o_{D-2}}\frac{\p \bar f(\bar r,\bar M,\bar \mu^i)}{\p\bar \mu^i}\Big|_{\bar r=\bar r_H}\d \bar \mu^i \label{1stkkd}
\end{align}
The generalized Smarr-like formula in arbitrary $D$ dimension is given as (see the appendix \ref{appen})
\begin{align}
(D-3)\bar M=(D-2)\bar T\bar S-\frac{\bar r_H^{D-2}}{\o_{D-2}}\frac{\p \bar f(\bar r,\bar M,\bar \mu^i)}{\p \bar r}\Big|_{\bar r=\bar r_H,\bar M=0}~.\label{smarrd}
\end{align}
Again, the last term in \eqref{smarrd} contributes when there is a source (or the cosmological constant). For, Schwartzschild metric in $D$-dimension we obtain the Smarr formula as $(D-3)\bar M=(D-2)\bar T\bar S$. Here, we have defined 
\begin{align}
\o_n=\frac{16\pi}{nVol(S^n)}=\frac{8\Gamma\Big(\frac{n+1}{2}\Big)}{n\pi^{\frac{n-1}{2}}}~,\label{omegan}
\end{align}
where $Vol(S^n)$ denotes the volume of a unit $n$-sphere $d\Omega_n^2$. In arbitrary dimension $D$, the expression of entropy is given as 
\begin{align}
\bar S=\frac{Vol(S^{D-2})\bar r_H^{D-2}}{4}~.
\end{align}
\vskip 5mm
\underline{\textit{RN-AdS black hole in $D$ spacetime dimensions:}}
\vskip 2mm
The spacetime metric of RN-AdS black hole in $D$ dimension is given by the $f(\bar r)$ as \cite{Cai:2001tv}
\begin{align}
\bar f(\bar r,\bar M,\bar \mu^i)=1-\frac{\o_{n}\bar M}{\bar r^{n-1}}+\frac{n\o_n^2\bar Q^2}{8(n-1)\bar r^{2n-2}}-\frac{2\bar \Lambda}{n(n+1)}\bar r^2~. \label{rnadsd}
\end{align}
Here $n=D-2$ corresponds to the number of angular coordinates. For this metric \eqref{rnadsd}, we obtain the first law from Eq. \eqref{1stkkd}, which is given by the Eq. \eqref{1STEXT1}. Thus, in the higher dimension, we get no change in the expression of the first law.

Here, $g(\bar r,\bar \mu^i)$ can be obtained as (see Eq. \eqref{frmmu} of the appendix \ref{appen})

Using the obtained Eq. \eqref{smarrd} we obtain the Smarr formula as
\begin{align}
(D-3)\bar M=(D-2)\bar T\bar S+(D-3)\bar Q\bar \Phi_H-2\bar P\bar V~, \label{smarrrnd}
\end{align}
which agrees with the existing results in literature \cite{Kastor:2009wy,Kubiznak:2016qmn}. However, as we have mentioned earlier, the importance of the present analysis lies in the fact that the results (1st law and the Smarr formula) can be obtained from the metric itself. In $D$-dimension, the potential ($\bar \Phi_H$) and the volume ($\bar V$) are defined as follows \cite{Cai:2001tv}.
\begin{align}
\bar V=\frac{Vol(S^n)}{n+1}\bar r_H^{n+1}~, \ \ \ \ \textrm{and} \ \ \ \ \bar \Phi_H=\frac{n}{4(n-1)}\frac{\o_n\bar Q}{\bar r_H^{n-1}}~.
\end{align}

Thus, in the present section, we have generally obtained the first law and the Smarr-like formula. Then we have considered a charged black hole in AdS space. We have found, if the cosmological constant is considered as a variable, we obtain the extra pressure and volume parameters in BH thermodynamics. Thus, the phase space of the black hole gets extended and the pressure is identified in terms of the cosmological constant and volume is its conjugate quantity. The mass of the black hole plays the role of enthalpy in the extended phase space. For RN black hole, the thermodynamic volume surprisingly matches to the geometric volume of a sphere with radius being the horizon radius. But, this may not be true in general. We shall highlight this later in more detail in our subsequent discussions. Another goal of our paper is to find the connection of thermodynamic parameters in two conformally connected frames. The thermodynamic parameters in GR is known. Therefore, in the following section, we define a conformal frame of GR and draw the connection of the parameters in the two frames.

Let us make some additional comments as follows. In our approach, the equilibrium state version first law and the general Smarr-like formula do not require any information regarding the conserved quantity. Here we obtain the first law and the Smaar formula in terms of the parameters which characterize the solution space of the BH under consideration. These parameters may not even arise from any conserved quantity. In GR, while obtaining first law for a static and spherically symmetric (SSS) black hole, the variation of the horizon radius is considered via two different viewpoints (i) physical process viewpoint, and (ii) equilibrium state version viewpoint. In the physical process view, one considers that the variation arises due to some physical process (such as the matter flux perturbing the horizon) and an SSS black hole evolves to another SSS black hole via this physical process. On the contrary, in the equilibrium state version viewpoint, one simply considers the variation arises due to the infinitesimal change in the parameters which characterizes the solution space of the BH. Actually, the equilibrium state version viewpoint compares the two nearby stationary solutions, which differ infinitesimally in terms of the parameters present in the solution space. Kindly note that our analysis is along the second route \textit{i.e} equilibrium state view. For example, the solution space of Einstein-scalar BH, as discussed in \cite{Lu:2014maa}, is characterized by the three parameters $\alpha$ (the mass parameter, which is not exactly the same as the AMD mass or the holographic mass), $\phi_1$, and $\phi_2$. None of these are the conserved quantities. The first law has been obtained in terms of these parameters (for example see Eq. (4.11) in \cite{Lu:2014maa}). Similarly, in our analysis, depending upon the source (or the energy-momentum tensor), several parameters ($\bar \mu^i$) will be present in the spacetime metric and the first law (or the Smarr formula) has been obtained in terms of those parameters ($\bar M$ and $\bar\mu^i$).

Note that, unlike the first law, the Smarr formula is non-universal. Even in our approach, we have found that the Smarr formula, unlike the first law, depends on the dimension of the spacetime. Here, we cannot claim that the procedure of obtaining the Smarr formula from the Einstein's equation is universal. In this paper, it has been found that for a general static and spherically symmetric spacetime (with $\bar g_{00}=-1/\bar g_{rr}$) the entropy of the BH (more precisely $\bar T\bar S$) is connected with the other parameters by a particular relation (kindly see Eq. \eqref{smarrf}/\eqref{smarrd}). We call this relation as the Smarr-like formula as it indicates similar insights as of the Smarr formula. In addition, we have found that, from this Smarr-like formula, one can obtain the Smarr formula for the well-known SSS black hole solutions such as the Schwartzschild BH, RN BH, RN BH in AdS space \textit{etc.} Therefore, at least in these cases, the Smarr formula can be obtained from the Einstein's equation.

 Note that the analysis presented in this section, will not be applicable for the SSS BH where $g_{00}\neq-1/g_{rr}$~. When the condition $\bar g_{00}=-1/\bar g_{rr}$ is relaxed, the analysis becomes more complicated. Therefore, we have kept is as the future work. Here, for a SSS metric with $\bar g_{00}=-1/\bar g_{rr}$~, we have shown the procedure to obtain the equilibrium state version of first law and the Smarr-like formula formula in a general way from the Einstein's equation. Our main goal, in this paper, is to study the extended phase space thermodynamics. We show that this general results provides the extended phase space thermodynamics, \textit{i.e.} the 1st law and the Smarr formula in any arbitrary dimension, by considering the specific example of the RNAdS black hole. Also, kindly note that when it has been claimed earlier to have obtained the first law from the Einstein's equation in a general way (see the refs. \cite{Padmanabhan:2002sha} and \cite{Hansen:2016gud}, where they have obtained the first law by defining thermodynamic potential), they also have considered $\bar g_{00}=-1/\bar g_{rr}$~. 

In order to obtain the extended phase space thermodynamics, here we have considered (following the standard prescriptions in the literature) the AdS spacetime, with the cosmological constant being a variable instead of being a constant. Thus, the extended phase space thermodynamics is a feature of AdS spacetime only. The question that whether there is analogue of thermodynamic pressure ($P$) in black hole spacetime, has been unsolved for long. In this regard, people have considered several quantities as the thermodynamic pressure in black hole spacetime. For example, in the earlier approach of obtaining the first law in a general form (for SSS spacetime) \cite{Padmanabhan:2002sha,Padmanabhan:2003gd}, they have considered $T^r_r$ (radial-radial component of the energy momentum tensor, or the radal pressure) as the thermodynamic pressure. This is a good consideration as the conjugate quantity of pressure naturally appears as the volume (of a sphere of horizon radius). However, this identification is not quite extensive as one cannot obtain other thermodynamic features (such as $P-V$ criticality) with this consideration. There are other interesting works in literature \cite{Chamblin:1999tk, Chamblin:1999hg}, which considers the inverse of Hawking temperature as thermodynamic pressure. With this identification, one can obtain the $P-V$ criticality. However, in that case, the expression of volume does not come naturally as the conjugate of thermodynamic pressure and also we do not have $PdV$ term in the first law.

Later, it was found that when cosmological constant is identified as pressure (\textit{i.e.} $\Lambda$ is considered as a variable instead of a constant, see the references \cite{Kastor:2009wy, Dolan:2010ha, Dolan:2011xt, Dolan:2011jm, Dolan:2012jh, Kubiznak:2012wp, Kubiznak:2016qmn,Majhi:2016txt,Bhattacharya:2017hfj,Bhattacharya:2017nru,Bhattacharya:2019qxe,Dehyadegari:2018pkb,Bhattacharya:2020jgk}), we naturally obtain the $PdV$ term in the first law as well as we obtain the $P-V$ criticality. Thus, in AdS spacetime (where $\Lambda$ is also considered as a variable) we have the proper analogue of thermodynamic pressure. This is called ``extended phase space thermodynamics'' (phase space is extended by the inclusion of $P$ and $V$). In asymptotically flat spacetime (or where $\Lambda=0$), we do not have this analogy between $P$ and $\Lambda$. Thus, in that case, we do not have proper analogue of thermodynamic pressure and volume and it is known as ``non-extended phase space thermodynamics''. In addition, $\Lambda=$constant case can be considered as a particular case of the extended phase space thermodynamics where $P=$constant.

Although the de Sitter (dS) spacetime is also characterized by the (positive) cosmological constant, analyzing black hole thermodynamics in   dS spacetime is much more complex. In de-Sitter spacetime, we have two different horizons, \textit{i.e.} black hole horizon along with cosmological horizon and both the horizons have different temperatures. As a result, the black hole thermodynamic system will be a non-equilibrium thermodynamic system. Thermodynamics of dS black hole system is different from standard thermodynamics in many ways. In addition, in dS spacetime we also have problem regarding defining the asymptotic mass. For these reasons, we have left the dS system. These issues (\textit{i.e.}, why we consider only AdS spacetime in extended phase space thermodynamics and not dS) has been discussed at length in the literature; for example, see \cite{Kubiznak:2016qmn}.
\section{Extended phase space thermodynamics in the conformally connected frame: study of (in)equivalence of thermodynamic parameters} \label{SECECON}
In the earlier section, we have shown that the Einstein's equation takes the form of the first law of black hole thermodynamics in the extended phase space. However, the van der Waals type phase transition (with the $P-V$ criticality) is also observed in Brans-Dicke theory, $f(R)$ theory \textit{etc.} which has been mentioned earlier. These theories can be described in either Jordan frame or in Einstein frame, which are conformally connected to each other. The physical equivalence of two conformal frames is a longstanding debate in literature \cite{Faraoni:1999hp, ALAL,Faraoni:1998qx, Faraoni:2010yi, Faraoni:2006fx, Saltas:2010ga, Capozziello:2010sc, Padilla:2012ze, Koga:1998un, Jacobson:1993pf, Kang:1996rj, Deser:2006gt, Dehghani:2006xt, Sheykhi:2009vc,Steinwachs:2011zs, Kamenshchik:2014waa, Banerjee:2016lco, Pandey:2016unk, Ruf:2017xon,Karam:2017zno,Bahamonde:2017kbs,Karam:2018squ,Bhattacharya:2017pqc,Bhattacharya:2018xlq,Bhattacharya:2020wdl,Bhattacharya:2020jgk,Dey:2021rke}. However, in our earlier work \cite{Koga:1998un,Bhattacharya:2017pqc,Bhattacharya:2018xlq,Dey:2021rke,Bhattacharya:2020jgk}, we have shown that the thermodynamic parameters, like entropy, temperature, internal energy (which is given by the mass of the black hole), angular momentum etc. are conformally invariant. But, the earlier analysis has been performed in the non-extended phase space. On the contrary, in the extended phase space, we have another two thermodynamic quantities: pressure ($P$) and volume ($V$)\footnote{Some recent works in the context of identification of $P$, $V$ and in the context of $P-V$ criticality can be found in \cite{Hendi:2017fxp,Hendi:2015hoa,Hendi:2015soe,Jafarzade:2017kin,EslamPanah:2019szt,Azreg-Ainou:2014lua,Azreg-Ainou:2014twa}.}. Earlier, these two parameters has not been defined in a covariant way for ST gravity and has not been investigated whether these are conformally equivalent (in \cite{Hyun:2017nkb}, the first law and the definition of $P$ and $V$ has been done in a covariant manner using the ADT formalism, but it is not obvious whether the same formalism will work in the present case). The conformal equivalence of other thermodynamic quantities in non-extended phase space do not guarantee the equivalence of $P$ and $V$ in the two conformally connected frames. Moreover, the thermodynamics in extended phase space is quite different compared to the non-extended one. The mass of black holes plays the role of internal energy in the non-extended phase space, whereas the same plays the role of enthalpy in the extended case. Therefore, in this section, we make a detail analysis for that. For simplicity, in the beginning, we take Einstein's gravity (GR) and later we compare it with a frame, which is conformally connected to GR. This is done because, the expressions of thermodynamic parameters in GR is already known and well-accepted. Therefore, we can straightforwardly compare the parameters of GR with the defined thermodynamic parameters in the conformal frame. In the following section, we shall generalize this result for the scalar-tensor theory in its two frame (\textit{i.e.} Jordan and the Einstein frame). The method, that we have adopted here, is the Noether charge formalism provided by Iyer and Wald \cite{Wald:1993nt,Iyer:1994ys}, which provides a powerful way to derive the first law and to define thermodynamic parameters in a covariant way. We do not incorporate the (energy-momentum) source term for simplicity. However, one can always incorporate source, which will contribute to the hairs of the black hole.

The Einstein-Hilbert action with the cosmological constant is given as 
\begin{align}
\bar{\mathcal{A}}=\int \sqrt{-\bar g}\bar L d^4x=\frac{1}{16\pi}\int \sqrt{-\bar g}(\bar R-2\bar\L) d^4x~. \label{AEH}
\end{align}

The procedure of obtaining the thermodynamic first law for the above action \eqref{AEH} (in the extended phase space) can be found in our earlier work \cite{Bhattacharya:2017hfj}. Therefore, instead of repeating it, we directly go to the conformal frame. We provide the following conformal transformation to the metric tensor $\bar g_{ab}$ and start the analysis in the conformal frame.
\begin{align}
\bar g_{ab}=\phi g_{ab}~, \label{CONF}
\end{align}
where $\phi$ is the conformal factor. In the conformal frame (described by the metric-tensor $g_{ab}$), one can obtain the equivalent action, which is given as 
\begin{align}
\mathcal{A}=\int \mg L=\frac{1}{16\pi}\int \mg \Big[\phi R+\frac{3}{2\phi}(\na_i\phi)(\na^i\phi)-3\square\phi-2\phi^{n}\L\Big]~, \label{ACTCON}
\end{align}
where $\L=\phi^{2-n}\bar\L$ (Note that here we have used $n$ as the exponent and not the spacetime index. This has been followed throughout). One can ask why we do not write the last term inside the square bracket of Eq. \eqref{ACTCON}  as $2\phi^2\bar\L$ instead of $2\phi^{n}\L$. For that, in the conformal frame, we define a variable $\Lambda$, which is the cosmological constant multiplied with the arbitrary power of the conformal factor (as $\L=\phi^{2-n}\bar\L$ and $n$ is kept arbitrary). Doing this, we shall keep the generality in the expression of the thermodynamic pressure as we shall go on to define $P=-\Lambda/8\pi$. The reason for keeping this generality is the following. In literature, there are several works in which the pressure has been accounted as the cosmological constant multiplied with the arbitrary power of the conformal factor. For example, in ref \cite{Chen:2013ce}, the thermodynamic pressure is identified as $P=-(\phi\bar\L)/8\pi$, which corresponds to the $n=1$ case of our analysis. Some other works \cite{Ovgun:2017bgx, Hendi:2015kza} accounts $P=\bar\L/8\pi$, which corresponds to $n=2$ of our analysis. Therefore, for a general analysis we keep it as $\phi^n\L$ in \eqref{ACTCON} and the pressure will be identified as $P=-\L/8\pi$. Later, we will check the consequences for different values of $n$.

Here we plan to obtain the covariant thermodynamic description using the Wald's formulation. For that, we need to obtain the Noether current and potential due to the diffeomorphism invariance. We obtain the Noether current and potential in the following.

The variation of the above action \eqref{ACTCON} (in the extended phase space approach \textit{i.e.} considering $\L$ as a variable) yields
\begin{align}
\d(\mg L)=\mg\Big[E_{ab}\d g^{ab}+E_{(\phi)}\d\phi-\frac{2\phi^{n}}{16\pi}\d\L+\na_a\Theta^a(q,\d q)\Big]~. \label{VARCON}
\end{align}
Here, $q=\{g_{ab},\phi\}$ and the expressions of $E_{ab}$, $E_{(\phi)}$ and $\theta^a(q, \delta q)$ are provided as follows:
\begin{eqnarray}
&& E_{ab}=\frac{1}{16\pi}\Big[\phi G_{ab}+g_{ab}\na_i\na^i\phi-\na_a\na_b\phi+\frac{3}{2\phi}(\na_a\phi)(\na_b\phi)-\frac{3}{4\phi}g_{ab}(\na^i\phi)(\na_i\phi)+g_{ab}\phi^{n}\Lambda\Big]~.
\no 
\\
&& E_{(\phi)}=\frac{1}{16\pi} \Big[R+\frac{3}{2\phi^2}(\na^i\phi)(\na_i\phi)-\frac{3}{\phi}\square\phi-2n \phi^{n-1}\Lambda\Big]~.
\no 
\\
&&\ \ \ \ \ \ \ \ \ \ \ \ \ \ \ \ \ \ \ \ \ \ \ \ \ \ \ \ \ \ \ \ \ \ \ \ \textrm{and}
\no 
\\
&& \theta^a(q, \d q)=\frac{1}{16\pi}\Big[2\phi P^{ibad}(\na_b\d g_{id})-2P^{iabd}(\na_b\phi)\delta g_{id}+\frac{3}{\phi}(\na^a\phi)\delta\phi-\frac{3}{2}g^{ij}(\na^a\phi)\d g_{ij}
\no 
\\
&&\ \ \ \ \ \ \ \ \ \ \ \ \ \ \ \ \ \ \ \ \ \ \ \ \ \ \ \ \ \ -3(\na_b\phi)\delta g^{ab}-3\na^a(\d\phi)\Big]~. \label{EXPLICITCON}
\end{eqnarray}
Here, we define $P^{abcd}$ as  $P^{abcd}=(g^{ac}g^{bd}-g^{ad}g^{bc})/2$. We can obtain a general Bianchi-like identity in this case as follows
\begin{align}
\na_b E^{ab}=-\frac{1}{2}E_{(\phi)}(\na^a\phi)+\frac{1}{16\pi}\phi^n(\na^a\Lambda)~. \label{BIANCHICON}
\end{align}
The above identity \eqref{BIANCHICON} will help us to obtain off-shell Noether current and potential due to the diffeomorphism invariance (the same identity, which is known as the generalized Bianchi identity or the Noether identity, also helps to obtain off-shell ADT current as well; for instance, see the ref \cite{Hyun:2016dvt}). Firstly, we want to obtain the change of the Lagrangian due to the diffeomorphism invariance $x^a\longrightarrow x^a+\xi^a$, which can be obtained by replacing arbitrary variation $\d$ in \eqref{VARCON} by Lie derivative ($\lie$). Thus, \eqref{VARCON} yields
\begin{align}
\na_a\Big(\mg L\xi^a\Big)=\mg\Big(-2E^{ab}(\na_a\xi_b)+E_{(\phi)}\xi^a\na_a\phi-\frac{2\phi^n}{16\pi}\xi^a\na_a\phi+\na_a\theta^a(q, \lie q)\Big)~. \label{preno}
\end{align}
To obtain Eq. \eqref{preno}, we have used $\lie g^{ab}=-(\na^a\xi^b+\na^b\xi^a)$, $\lie L=\xi^a\na_a L$, $\lie\phi=\xi^a\na_a\phi$, $\lie\L=\xi^a\na_a\L$ and so on. In \eqref{preno}, we use Bianchi like identity (given in \eqref{BIANCHICON}). This makes the RHS of \eqref{preno} a total derivative term and one can show \eqref{preno}, as a whole, can be written as $\na_a J^a=0$, where the conserved off-shell Noether current $J^a$ is given as 
\begin{align}
J^a=L\xi^a+2E^{ab}\xi_b-\theta^a(q, \lie q)~. \label{NCURRENTCON}
\end{align}
The above Noether current \eqref{NCURRENTCON} can be further expressed as a two-ranked anti-symmetric potential, also known as the Noether potential, which is given by the relation $J^a=\na_bJ^{ab}$\footnote{see appendix A of \cite{Bhattacharya:2018xlq} and consider $\o(\phi)=-3/2$ and $V(\phi)=2\phi^n\L$ which mimics our system~.}, where 
\begin{align}
J^{ab}[\xi]=\frac{1}{16\pi}\Big[\na^a(\phi\xi^b)-\na^b(\phi\xi^a)\Big]~. \label{NOPCON}
\end{align}
So far, the Noether current and the potential is obtained off-shell, where the equation of motion is nor required to be used. In addition, the above expressions of the Noether current and potential are obtained for any arbitrary diffeomorphism. We shall follow the Wald's formulation to obtain the thermodynamic law in the extended phase space. In that case, it required to take the on-shell variation (or perturbation) of the Noether current with respect to the fields $g_{ab}$ and $\phi$ (therefore, $\d\xi^a=0$ but $\d\xi_a\neq 0$). In order to adopt Wald's formalism, we need to consider the arbitrary diffeomorphism as the Killing diffeomorphism. From here onward, $\xi^a$ will be considered as the Killing vector. The onshell perturbation of the Noether current is given as
\begin{align}
\d(\mg J^a)=\xi^a\d(\mg L)-\d(\mg\theta^a(q, \lie q))
\no 
\\
=\mg\xi^a\na_b\theta^b(q, \d q)-\d(\mg\theta^a(q, \lie q)) -\mg\frac{2\phi^n}{16\pi}\xi^a\d\Lambda~. \label{VARJCON}
\end{align}
Upon using $\mg\xi^a\na_b\theta^b(q, \d q)=\lie(\mg \theta^a(q, \d q))+2\mg\na_b\xi^{[a}\theta^{b]}(q, \d q)$ in the above equation \eqref{VARJCON} (where $A^{[a}B^{b]}=(A^aB^b-A^bB^a)/2$), one obtains
\begin{align}
\d(\mg J^a)=-\boldsymbol\o^a+2\mg\na_b\xi^{[a}\theta^{b]}(q, \d q)-\mg\frac{2\phi^n}{16\pi}\xi^a\d\Lambda~,
\end{align}
where,
\begin{align}
\boldsymbol\o^a=-\lie(\mg \theta^a(q, \d q))+\d(\mg\theta^a(q, \lie q))
\end{align}
is identified as the symplectic Hamiltonian (for rigourous discussion see \cite{Wald:1993nt}). Variation of the total Hamiltonian is given as 
\begin{align}
\d H=\int_c d\Sigma_a\frac{\boldsymbol\o^a}{\mg}=-\delta\int_c d\Sigma_a\na_bJ^{ab}+2\int_c d\Sigma_a \na_b\xi^{[a}\theta^{b]}(q, \d q)-\frac{\d\Lambda}{8\pi}\int_c\phi^nd\Sigma_a\xi^a~. \label{DELHCON}
\end{align}
The above integration is performed on a three-dimensional Cauchy hypersurface, which has been symbolized by $``c"$ in the subscript of the integral. Here, $d\Sigma_a=n_a\sqrt{h}d^3x$ is the elemental surface area of the hypersurface, with $n_a$ being the normal and $h$ being the determinant of the induced metric of the surface. The first two terms of RHS of Eq. \eqref{DELHCON} can be written in terms of two-surface integral applying the Stoke's theorem. Since the spacetime contains a horizon, the two surface will not be a closed one. Instead, it will have two ends-- one is the black hole horizon ($\mathcal{H}$) and the other one is the asymptotic infinity ($\p c_{\infty}$). Following Wald \cite{Wald:1993nt,Iyer:1994ys}, we also consider that the black hole horizon is a bifurcation Killing horizon, where the Killing normal $\xi^a=0$. One can check, for the presence of the Killing vector in the spacetime, $\boldsymbol\o^a$ vanishes (\textit{i.e.} $H$ vanishes). Thereby, we obtain 
\begin{align}
-\frac{1}{2}\d\int_{\mathcal{H}}d\Sigma_{ab}J^{ab}[\xi]+\frac{1}{2}\d\int_{\p c_{\infty}}d\Sigma_{ab}J^{ab}[\xi]-\frac{1}{16\pi}\int_{\p c_{\infty}}d\Sigma_{ab}\xi^{[a}\theta^{b]}(q, \d q)+\d P\int_c\phi^nd\Sigma_a\xi^a=0~, \label{prefirst}
\end{align}
where we identify $P=-\Lambda/8\pi$ and $d\Sigma_{ab}=\sqrt{\sigma}(\xi_al_b-\xi_bl_a)$ is the elemental surface area of the two surfaces ($\mathcal{H}$ and $\p c_{\infty}$), with $\boldsymbol{l}$ being the auxiliary null vector and $\sigma$ being the determinant of the induced metric on the two surface. We consider the black hole metric to be stationary-axisymmetric, which possess the Killing vector $\boldsymbol{\xi}=\boldsymbol{\xi_{(t)}}+\Omega\boldsymbol{\xi_{\phi}}$, where $\boldsymbol{\xi_{(t)}}$ and $\boldsymbol{\xi_{\phi}}$ are the components of the Killing vector along the temporal and the azimuthal directions respectively and $\Omega$ is the angular velocity. Then the above relation \eqref{prefirst} can be identified as the first law in the extended phase space, which is given as
\begin{align}
\delta M=T\d S+\Omega_H\delta J+ V\d P~, \label{1stlawcon}
\end{align}
where, following Wald's approach, one can identify mass ($M$), entropy ($S$) and the angular momentum ($J$) as
\begin{eqnarray}
&& \d S=\frac{\pi}{\kappa}\delta\int_{\mathcal{H}} d\Sigma_{ab}J^{ab}~;
\no 
\\
&& \d M=\frac{1}{2}\int_{\partial c_{\infty}} [\d(d\Sigma_{ab}J^{ab})-2d\Sigma_{ab}\xi^{[a}\T^{b]}(q, \d q)]\Big|_{\xi=\xi_{(t)}}~;
\no 
\\
&& \d J=-\frac{1}{2}\int_{\partial c_{\infty}} [\d(d\Sigma_{ab}J^{ab})-2d\Sigma_{ab}\xi^{[a}\T^{b]}(q, \d q)]\Big|_{\xi=\xi_{(\phi)}}~.
 \label{SMJ}
\end{eqnarray}
In the above relations \eqref{SMJ}, $\kappa$ is the surface gravity of the Killing horizon and the temperature is identified as $T=\kappa/2\pi$. 

Let us now comment upon the thermodynamic volume (denoted as $V$) in the following. A lot of work has been done \cite{Caceres:2015vsa, Karch:2015rpa, Caceres:2016xjz, Nguyen:2015wfa, Kastor:2016bph, Kastor:2014dra,Pradhan:2016rff,Couch:2016exn} in the context of thermodynamic volume in the extended phase space. In general, thermodynamic volume, is the conjugate quantity of $P$ and may not be the same as the geometric volume. The expression of the thermodynamic volume is provided in the following way. From Eq. \eqref{prefirst}, we obtain the conjugate quantity of $P$ as
\begin{align}
V=\int_c \phi^n d\Sigma_a\xi^a~.
\end{align}
Here $d\Sigma_a$ corresponds to the elemental surface area of the Cauchy surface $c$ (also note that $n$ is the exponent of $\phi$ and not the indices). Now, in several cases (see the examples given in \cite{Kastor:2009wy,Couch:2016exn}), it has been found that, with this definition, $V$ appears as the divergent quantity when it is integrated in the whole space (of course those analysis in \cite{Kastor:2009wy,Couch:2016exn} are for Einstein's gravity, where $\phi=1$). A careful investigation suggests that this divergence mainly appears due to the presence of the AdS part (see \cite{Couch:2016exn}). \textit{i.e.} $\int_c  d\Sigma_a\xi^a$ is a divergent quantity for a pure AdS spacetime where black hole is not present. Therefore, this pure AdS contribution part is removed from the definition of the thermodynamic volume. This is done in order to obtain a finite expression of thermodynamic volume. This is called the background subtraction. Therefore, the regularized expression of thermodynamic volume is given as
\begin{align}
V=\int_c \phi^n d\Sigma_a\xi^a-\int_c (\phi^n d\Sigma_a\xi^a)_{AdS}~. \label{volvol1}
\end{align}

The above expression of thermodynamic volume can be expressed in another form, which resembles more with the expressions given in the literature. Since $\xi^a$ is a Killing vector, it can be proved  that $\phi^n\xi^a$ is divergenceless (\textit{i.e.} $\nabla_a(\phi^n\xi^a)=0$). Therefore, we can introduce a two-ranked anti-symmetric super-potential $\chi^{ab}$ as \cite{Kastor:2009wy}
\begin{align}
\phi^n\xi^a=\nabla_b(\chi^{ab})
\end{align}
The existence of $\chi^{ab}$ is guaranteed, at least locally, by the Poincare's lemma. Therefore, applying Stoke's theorem, the above expression of thermodynamic volume can be expressed in terms of the two-surface integration, \textit{i.e.}
\begin{align}
V=\frac{1}{2}\Big[\int_{\partial c}d\Sigma_{ab}\chi^{ab}-\int_{\partial c}(d\Sigma_{ab}\chi^{ab})_{AdS}\Big] \label{prevol}
\end{align}
This two-surface (which is denoted by $\partial c$) has two ends, one at the horizon (denoted by $\mathcal{H}$) and the other one is asymptotic infinity (denoted by $\partial c\infty$). Thus, we finally obtain the expression of the thermodynamic volume as
\begin{align}
V=\frac{1}{2}\Big[\int_{\mathcal{H}}d\Sigma_{ab}\chi^{ab}-\int_{\p c_{\infty}}d\Sigma_{ab}\Big(\chi^{ab}-\chi^{ab}_{(AdS)}\Big)\Big], \label{vol}
\end{align}
Since $\chi^{ab}_{AdS}$ is the contribution arising from pure AdS spacetime, it does not have any contribution on the black hole horizon. This expression of volume has been provided in the equation (51), which the referee has mentioned. Also, this expression of thermodynamic volume resembles to the expression, obtained in earlier literature, such as ref. \cite{Kastor:2009wy}. Moreover, as we see in the expression of the first law, as provided in Eq \eqref{1stlawcon}, the mass of the black hole plays the role of enthalpy instead of internal energy. Furthermore, the Wald's method is applicable only when there is a Killing symmetry in the spacetime and the BH horizon is the Killing horizon. However, for the two conformally connected frames, if there is a Killing vector in one frame, it will be a conformal Killing vector in the other frame and the BH horizon will be the conformal Killing horizon. However, if the conformal factor (here $\phi$) is Lie-transported along the direction of the Killing vector (\textit{i.e.} $\lie\phi=0$), the Killing vector in one frame will be the Killing vector in the other frame. Thus, in our analysis, we have considered $\phi$ to be Lie-transported along the direction of the Killing vector. This is a standard practice and have also been adopted in earlier works (such as \cite{Koga:1998un}).

\subsection*{Equivalence of the thermodynamic parameters}
Let us now see how the thermodynamic parameters of Einstein's gravity and the same in the conformally connected frames are related. As we have mentioned, using the same Iyer-Wald technique, the first law has been proved for Einstein-Hilbert action and the thermodynamic parameters have also been obtained \cite{Bhattacharya:2017hfj}. In Einstein's gravity, the 1st law and the expression of the thermodynamic parameters are provided as follows.
\begin{align}
\d\bar M=\bar T\d\bar S+\bar\Omega_H\delta\bar J+ \bar V\d\bar P~, \label{1stlaweh}
\end{align}
where the thermodynamic parameters are provided as 
\begin{eqnarray}
&& \d \bar S=\frac{\pi}{\bar \kappa}\delta\int_{\mathcal{H}} d\bar \Sigma_{ab}\bar J^{ab}~;
\no 
\\
&& \d\bar  M=\frac{1}{2}\int_{\partial c_{\infty}} [\d(d\bar \Sigma_{ab}\bar J^{ab})-2d\bar \Sigma_{ab}\bar \xi^{[a}\bar \T^{b]}(\bar q, \d \bar q)]\Big|_{\bar \xi=\bar \xi_{(t)}}~;
\no 
\\
&& \d \bar J=-\frac{1}{2}\int_{\partial c_{\infty}} [\d(d\bar \Sigma_{ab}\bar J^{ab})-2d\bar \Sigma_{ab}\bar \xi^{[a}\bar \T^{b]}(\bar q, \d\bar q)]\Big|_{\bar \xi=\bar \xi_{(\phi)}}~.
 \label{SMJeh}
\end{eqnarray}
In the above relation, $q$ corresponds to the metric tensor describing the Einstein's gravity in its original frame (\textit{i.e.} $q=\bar g_{ab}$). For GR, the Noether potential is given as $\bar J^{ab}=(\bar\na^a\bar\xi^b-\bar\na_b\bar\xi^a)/16\pi$ and the boundary term is given as $\bar \T^{a}(\bar q, \d\bar q)=2\bar P^{ibad}\bar\nabla_b\d \bar g_{id}$ \cite{paddybook}. Here, we have assumed that the Killing vectors in the two frames coincide (\textit{i.e.} $\bar\xi^a=\xi^a$, implying $\bar\xi_a=\phi\xi_a$). Thus, the surface gravity $\kappa=\sqrt{((\na_a\xi^b)(\na_b\xi^a))/2}$ and $\bar \kappa=\sqrt{((\bar \na_a\bar \xi^b)(\bar \na_b\bar \xi^a))/2}$ can be proved to be the same on the bifurcation 2-surface (where $\xi^a=\bar\xi^a=0$). The proof of the zeroth law \cite{Dey:2021rke} suggests that the constancy of $\kappa$ (or $\bar\kappa$) holds not only on the bifurcation surface, but also all over the BH horizon. Thus, $\kappa=\bar\kappa$ is valid all over the BH horizon. Which implies $\bar T=T$. The equivalence of the angular velocity can be proved along the same line of argument provided in \cite{Koga:1998un}. One can prove that the Noether potential and the surface term are related in the two frames as 
\begin{align}
\bar J^{ab}=\frac{J^{ab}}{\phi^2}
\end{align}
and 
\begin{align}
\bar \T^{a}(\bar q, \d\bar q)=\frac{\T^{a}(q, \d q)}{\phi^2}~.
\end{align}
One can obtain the elemental surface area of the two surface is connected in the two frames as $d\bar\Sigma_{ab}=\phi^2d\Sigma_{ab}$. Thus, one can prove the equivalence of the thermodynamic parameters as $\bar M=M$, $\bar S=S$ and $\bar J=J$ from the two equations \eqref{SMJ} and \eqref{SMJeh}. in the action \eqref{ACTCON}, the term $3\square\phi$ is a total derivative term, which could have been discarded from the action as the removal of a total derivative term does not affect the dynamics of the fields. However, if we had discarded $3\square\phi$ term, the exact equivalence of mass, angular-momentum and entropy could not have been established (for more discussion along this line, please see \cite{Bhattacharya:2018xlq,Bhattacharya:2020jgk}). Thus, the $3\square\phi$ is very important in proving the exact equivalence of the thermodynamic parameters.

Let us discuss upon the equivalence of pressure and volume. As we have identified $\Lambda=\phi^{2-n}\bar\Lambda$, we obtain 
\begin{align}
P=\phi^{2-n}\bar P~.
\end{align}
Thus the expression of pressure is not equivalent in the two frames unless $n=2$. The expression of volume is also equivalent iff $n=2$. Thus, we can say, in the conformal frame for Einstein-Hilbert action, the thermodynamic quantities, such as mass, entropy, angular-momentum, temperature \textit{etc.} are equivalent in the two frames and the equivalence of thermodynamic pressure and volume depends upon the choice of $n$. However, the first law can be proved for any arbitrary choice of $n$. Thus our analysis does not guarantee that the pressure and volume has to be conformally equivalent. Since the other parameters (such as mass, entropy \textit{etc.}) are explicitly shown to be equivalent in the two frames, one can expect the same. The formalism, mentioned above, is applicable irrespective of the equivalence of pressure and volume (\textit{i.e.} irrespective of the choice of $n$). 

In the next section, we show that the same results are valid for a more general case, \textit{i.e.} for the scalar-tensor gravity which is described in the Jordan and the Einstein frame, which are conformally connected. The analysis presented in this section will provide us the hint for the general ST gravity, which is presented in the following section. 
\section{Extended phase space thermodynamics in the scalar-tensor gravity} \label{SECST}
In this section, we generalize our earlier results for the scalar-tensor gravity, which is described in two frames. In the original frame, which is known as the Jordan frame, the action is given as
\begin{eqnarray}
&&\mathcal{A}^{(ST)}=\int d^4x\sqrt{-g} \frac{1}{16\pi}\Big(\phi R-\frac{\omega (\phi)}{\phi}g^{ab}\nabla_a\phi \nabla_b\phi -\V(\phi)\Big)~.
\label{SJ}
\end{eqnarray}
In this frame, the scalar field $\phi$ is nonminimally coupled with the Ricci-scalar $R$. In addition, $\omega(\phi)$ is known as the Brans-Dicke parameter, which is considered as an arbitrary function of the scalar field $\phi$. When $\omega(\phi)$ is constant, this theory boils down to the Brans-Dicke (BD) theory. Furthermore, we have the scalar field potential $\V(\phi)$, which can also contain the cosmological constant $\bar\Lambda$ (see ref \cite{Hendi:2015kza,Hendi:2015hgg} for instance), in extended phase space approach which will be regarded as a variable. Thus, from now onward, we consider $\V(\phi)\equiv \V(\phi,\bar\Lambda)$. The nonminimal coupling in the original frame can be removed by a set of transformation, one is the conformal transformation and the other one is the rescaling of the scalar field. The transformation relations are given as
\begin{align}
\tilde{g}_{ab}=\phi g_{ab},
\label{GAB}
\end{align}
and
\begin{align}
 \phi\rightarrow\tilde{\phi}\,\ {\textrm{with}}\,\ d\tilde{\phi}=\sqrt{\frac{2\omega(\phi)+3}{16\pi}}\frac{d\phi}{\phi}~.
\label{PHI}
\end{align}
With the help of above two transformation relations \eqref{GAB} and \eqref{PHI}, one can write the equivalent form of the action \eqref{SJ} in the Einstein frame as
\begin{eqnarray}
\tilde{\mathcal{A}}=\int d^4x\sqrt{-\t g}\t {\mathcal{L}}=\int d^4x\sqrt{-\tilde{g}}\Big[\frac{\tilde{R}}{16\pi}-\frac{1}{2}\tilde{g}^{ab}\tilde{\nabla}_a\tilde{\phi}\tilde{\nabla}_b\tilde{\phi}-\U(\tilde{\phi}, \bar\Lambda)\Big]~,
\label{SE}
\end{eqnarray}
where $\U(\tilde{\phi},\bar\Lambda) = \V(\phi,\bar\Lambda)/(16\pi\phi^2)$.
In literature, one finds that the action of ST gravity in the two frames are provided by Eqs \eqref{SJ} and \eqref{SE}. However, the two actions \eqref{SJ} and \eqref{SE} are not exactly equivalent under the transformation relations \eqref{GAB} and \eqref{PHI}. Instead the two actions (or the Lagrangians) differ by a total derivative term ($-3\square\phi$). If one neglects $-3\square\phi$ term in the action the dynamics of the system remains invariant but, there remains an in-built inequivalence in the action, which results in the several inequivalences in the two frame; such as the violation of holographic principal in the Jordan frame, inequivalence of Noether current and potential in the two frames, which further lead to inequivalence of the thermodynamic parameters \textit{etc.} (for more discussions in this regard, see \cite{Bhattacharya:2017pqc,Bhattacharya:2020jgk}). For the specific choice of $\o(\phi)=-3/2$ and $\U(\tilde{\phi},\bar\Lambda) =\bar\Lambda/8\pi$, one recovers the Einstein's gravity with the cosmological constant term (where Eq. \eqref{SE} boils down to the Eq. \eqref{AEH}). As we have mentioned earlier, in several works, the thermodynamic pressure for $f(R)$ and BD theory has been considered as the cosmological constant multiplied with different powers of the the conformal factor (here $\phi$) in different cases. Therefore, to maintain the generality in the expression of pressure, we consider the thermodynamic pressure in the Jordan  frame is given as $P=-\Lambda/\pi$. Whereas, in Einstein frame we defined it as $\t P=-\bar\Lambda/8\pi$, where $\Lambda$ and $\bar\Lambda$ are related by   $\Lambda=\phi^{(n-2)}\Lambda$\footnote{The thermodynamic pressure in the Einstein frame is considered as the same of GR due to the fact that in the limit $\o(\phi)=-3/2$, Einstein frame action boils down to the Einstein-Hilbert action.}. Thus, incorporating $-3\square\phi$ term and considering $\Lambda$ as a variable instead of $\bar\Lambda$, the action of the Jordan frame is given as 
\begin{align}
\mathcal{A}=\int d^4x\sqrt{-g}\mathcal{L} =\int d^4x\sqrt{-g} \frac{1}{16\pi}\Big(\phi R-\frac{\omega (\phi)}{\phi}g^{ab}\nabla_a\phi \nabla_b\phi -\V(\phi, \Lambda)-3\square\phi\Big)~.\label{SJNEW}
\end{align}
From now onward, the action of the Jordan frame will be considered as $\mathcal{A}$ (given in \eqref{SJNEW}) instead of $\mathcal{A}^{(ST)}$ (given in \eqref{SJ}) for the aforementioned reasons.

In both the frames, one can obtain the first law following the Wald's approach as shown in the previous section and can define the thermodynamic parameters as earlier. In order to avoid repeating similar calculations, in appendix \ref{AppenST} we have obtained the first law and have shown that Mass, entropy, angular momentum, angular velocity and temperature (\textit{i.e.} the non-extended phase space variables) are conformally equivalent \footnote{In the context of ST gravity, the non-extended phase space variables are already defined and have shown equivalent in the two frames \cite{Koga:1998un,Bhattacharya:2017pqc,Bhattacharya:2018xlq,Dey:2021rke,Bhattacharya:2020jgk}. However, the present discussion is valid for both non-extended phase space variables as well as extra variables ($P$ and $V$) appearing in the extended phase space but, here we emphasize on $P$ and $V$ only.}. The pressure in the Jordan frame is already identified as $P=-\L/8\pi$ and, the thermodynamic volume is identified as 

\begin{align} 
  V=-\frac{1}{2}\int_{\mathcal{H}} \frac{\p \V(\phi,\L)}{\p\L}\sqrt{h} n_a\xi^ad^3y+\frac{1}{2}\int_{\mathcal{\infty}}\frac{\p \V(\phi,\L)}{\p\L} \sqrt{h} \Big[n_a\xi^a-(n_a\xi^a)_{BG}\Big]d^3y~, \label{VOLJOR}
\end{align}  
where $n_a$ is the normal and $h$ is the determinant of the induced metric of $t=$ const. hypersurface\footnote{the above expression of volume \eqref{VOLJOR} can be expressed as the integration of two surface (introducing Killing-superpotential, as done in the section \ref{SECECON}) when $\V(\phi,\L)$ is $\mathcal{O}(\L)$, as a result of which $\p V(\phi,\L)/\p\L$ is a function of $\phi$ only. For example, it can be done for the $\V(\phi,\L)$ given in \cite{Hendi:2015kza} (Eq. 13).}

Similarly, in Einstein frame, the expression of pressure is given as $\t P=-\bar\L/8\pi$ and the expression of thermodynamic volume is given as 

\begin{align} 
  \t V=-8\pi\int_{\mathcal{H}} \frac{\p \U(\t \phi,\bar \L)}{\p\bar \L}\sqrt{\t h} \t n_a\t\xi^ad^3y+ 8\pi\int_{\mathcal{\infty}}\frac{\p \U(\t \phi,\bar \L)}{\p\bar \L}\sqrt{\t h} \Big[\t n_a\t\xi^a-(\t n_a\t\xi^a)_{BG}\Big]d^3y~, \label{VOLEIN}
\end{align} 
Again, here we have adopted the regularization prescription in order to ward off the divergence appearing in the expressions of volume as provided in \eqref{VOLJOR} and \eqref{VOLEIN}.
One can easily check and find that the thermodynamic pressure and volume will be equivalent in the two frames only when $n=2$ and $\L=\bar\L$ (thus $P=\bar P=-\bar\L/8\pi$). 

We provide some important comments about the analysis done above. As we have mentioned earlier, it has been found that the expression of thermodynamic pressure for alternative theories of gravity, such as Brans-Dicke theory, $f(R)$ theory etc., have been considered different in different literature while studying the $P-V$ criticality. Whereas, the expression of thermodynamic pressure in Einstein's gravity is unanimously accepted. The $f(R)$ theory and the Brans-Dicke theory can be studied as the special cases scalar-tensor gravity in the Jordan frame. On the other hand, one can roughly say that the GR can be considered as the special case of scalar-tensor theory in the Einstein frame. It has been found that the thermodynamic parameters of non-extended phase space are equivalent in the two conformally connected frames \cite{Koga:1998un,Bhattacharya:2017pqc,Bhattacharya:2018xlq,Dey:2021rke,Bhattacharya:2020jgk}. Therefore, normally it can be expected that the extra two parameters in the extended phase space approach (\textit{i.e.} $P$ and $V$) should also be equivalent in the two frames. Nevertheless, the formalism described above to obtain the first law is valid irrespective of the equivalence of pressure and volume.

Thus, in our analysis, we have first considered the Einstein's gravity (GR) and have compared the thermodynamic parameters of GR (in the extended phase space approach) with that of its conformal frame. Here, we have adopted Wald's approach, which have been found useful not only to prove the first law but also to provide the expression of the thermodynamic parameters in a covariant manner and, thereby, establishing connection of the thermodynamic parameters in the two frames. Our analysis suggests, if thermodynamic pressure and volume have to be equivalent in the two frames (which is most likely as other thermodynamic parameters are shown to be equivalent) one always have to consider thermodynamic pressure in terms of cosmological constant only. Later, we have generalized this result for scalar-tensor gravity. The significance in our approach lies in the fact that we have provided all the expression of the thermodynamic parameters in a covariant way. 

\section{Conclusions and outlook} \label{SECCON}
Extended phase space thermodynamics has been studied extensively under the framework of Einstein's GR, where the cosmological constant is considered as the thermodynamic pressure ($P$) and its conjugate quantity is considered as the thermodynamic volume ($V$). However, recently the $P-V$ criticality has also been found for Brans-Dicke and for the $f(R)$ gravity, which can be studied as a special case of the ST gravity and can be described in the two conformally connected frames. However, there remains a discrepancy in the expression of pressure in different literature. In some literature, the cosmological constant alone is considered as the pressure whereas, in another case, the cosmological constant, multiplied with conformal factor is identified as the thermodynamic pressure. In addition, the equivalence of the thermodynamic parameters, in the extended phase space approach, has not been studied for the conformally connected frames.

Motivated by these facts, we present the analysis of this paper. Firstly in GR, we provide a general way of obtaining the first law and the Smarr formula from Einstein's equation, which works both for the extended phase space and for the non-extended phase space. Generally, while proving the first law from the Einstein's equation, people define thermodynamic potential and obtain the equilibrium version of first law in terms of the virtual change of the  defined thermodynamic potential. However, originally the equilibrium version of the first law is obtained in terms of the virtual change of the parameters present in the metric. We provide a general analysis along this route. Thereby we also show that in extended phase space thermodynamics, the pressure can be identified in terms of the cosmological constant. In addition, we show that the Einstein's equation contains more information on BH thermodynamics as one can obtain the Smarr formula from the Einstein's equation itself, which was not predicted earlier. This result is significant in the context of studying gravity as an emergent phenomena. Later, we investigate the connection of thermodynamic parameters in the two conformally connected frames. Since, the thermodynamic parameters in Einstein's GR is already well-known and well-accepted, we introduce the conformal frame of GR and obtain the first law using Iyer-Wald formalism. Thereby, we also obtain the expression of thermodynamic parameters of extended phase space in a covariant manner, which we compare with the parameters of Einstein's GR. We find, in the conformal frame, the equivalence of thermodynamic parameters is attainable when the pressure is identified in terms of cosmological constant only and not multiplied with the conformal factor. Finally, we have extended this analysis for the general scalar-tensor gravity in its two frames, where the similar conclusions are drawn.

Here we discuss one problematic issue related to this topic. The extended phase space thermodynamics seems to be incompatible with AdS-CFT correspondence. As per AdS-CFT, the thermodynamics of AdS black hole can be described equivalently by the dual CFT. Several earlier works suggest that the variation of  cosmological constant, which is identified as the thermodynamic pressure in the bulk, is not dual to the pressure of the dual field theory. Instead, it has been argued \cite{Kastor:2009wy,Johnson:2014yja,Dolan:2014cja,Kastor:2014dra,Caceres:2015vsa,Karch:2015rpa} that varying cosmological constant is dual to varying the number of colours ($N$) in the dual gauge theory. If the gauge theory is arising from coincident D-branes, varying $N$ corresponds to the number of branes. Note that $N$ is a discrete number and not a thermodynamic variable. In addition, thermodynamics is about the dynamics of the microstates at the equilibrium. However, the variation of the cosmological constant (or $N$) corresponds to the change in theory itself and not just the change in state. Here, we just acknowledge this issue as we are not describing the extended phase space thermodynamics from AdS-CFT viewpoint. Furthermore, despite this issue, the extended phase space thermodynamics has been a subject of ongoing research (the reference of this paper lists a few of them). Also, see the new works \cite{Visser:2021eqk,Cong:2021jgb}, where these issues are discussed. Hopefully, this issue will be resolved in future works.

 Thus, in this work, the extended phase space thermodynamics has been thoroughly studied for two conformally connected frames to address the discrepancy in the literature. Most importantly, we have provided the expressions of thermodynamic parameters in a covariant way following standard existing formalisms of GR. In addition, we have shown the procedure to obtain equilibrium version of first law and the Smarr-like formula in a general way for a static and spherically symmetric black hole in GR. We hope the analysis provides some significant insights in the context of extended phase space thermodynamics. More on this topic is expected to be reported soon.

\vskip 3mm
\noindent
{\bf Acknowledgement:} I am thankful to Late Prof. Thanu Padmanabhan for his comments on the draft of this article. Special thanks to Prof. Bibhas Ranjan Majhi for his comments on this article. At the communication stage, KB moved to IACS (Kolkata) and later Fukushima University (JSPS fellow). Therefore, KB also acknowledges the financial assistance received from IACS, and from JSPS during the time of communication.

  \vskip 3mm
 \textbf{Data availability:} This manuscript has no associated data or the data will not be deposited.

\appendix
\section{Obtaining the first law and the Smarr-like formula in arbitrary dimension} \label{appen}
For the metric \eqref{metricd}, the Einstein's equation is given as
\begin{align}
(D-3)\Big(-1+\bar f(\bar r)\Big)+\bar r\bar f'(\bar r)=\frac{16\pi}{D-2}\bar r^2\bar T^{\bar r}_{\bar r}~.
\end{align}
The general solution of the above equation comes in the following form
\begin{align}
\bar f(\bar r,\bar \mu^i)=1-\Big(\frac{\bar r_H}{\bar r}\Big)^{D-3}+\frac{1}{\bar r^{D-3}}\int_{\bar r_H}^{\bar r}\frac{16\pi}{D-2}T^{\bar r}_{\bar r}(\bar r,\bar \mu^i)\bar r^{D-2}d\bar r~.
\end{align}
This expression can be written in the following form
\begin{align}
\bar f(\bar r,\bar \mu^i)=g(\bar r,\bar \mu^i)-\Big(\frac{\bar r_H}{\bar r}\Big)^{D-3}g(\bar r_H,\bar \mu^i)~.\label{frmu}
\end{align}
We identify 
\begin{align}
\omega_{D-2}\bar M=\bar r_H^{D-3}g(\bar r_H,\bar \mu^i)~,\label{omegam}
\end{align}
and, thereby, we can write Eq. \eqref{frmu} as
\begin{align}
\bar f(\bar r,\bar M,\bar \mu^i)=g(\bar r,\bar \mu^i)-\frac{\omega_{D-2}\bar M}{\bar r^{D-3}}~,\label{frmmu}
\end{align}
where 
\begin{align}
g(x,\bar \mu^i)=1-\frac{1}{x^{D-3}}\int^x\frac{2}{D-2}\epsilon(x,\bar \mu^i)x^{D-2}dx~.\label{gxmu}
\end{align}
Eq. \eqref{frmmu} implies that $g(\bar r,\bar \mu^i)=\bar f(\bar r,\bar M,\bar \mu^i)_{\bar M=0}$~. From Eq. \eqref{frmu}, we obtain
\begin{align}
\bar f'(\bar r_H, \bar M,\bar \mu^i)=g'(\bar r_H,\bar \mu^i)+\frac{D-3}{\bar r_H}g(\bar r_H,\bar \mu^i)~.\label{fprimemmu}
\end{align}
Replacing $g(\bar r_H,\bar \mu^i)$ of Eq. \eqref{omegam} using the above Eq. \eqref{fprimemmu} we obtain Eq. \eqref{smarrd}.
Let us now consider the horizon radius changes infinitesimally due to the change in parameters ($\bar M$ and $\bar \mu^i$). In that case we have (from Eq. \eqref{omegam})
\begin{align}
\o_{D-2}\d \bar M=\Big[(D-3)\bar r_H^{D-4}g(\bar r_H,\bar \mu^i)+\bar r_H^{D-3}g'(\bar r_H,\bar \mu^i)\Big]\d \bar r_H+\bar r_H^{D-3}\frac{\p g(\bar r_H,\bar \mu^i)}{\p \bar mu^i}\d\bar \mu^i~.\label{omegadelm}
\end{align}
Using equations \eqref{gmu} and \eqref{fprimemmu} in \eqref{omegadelm}, we obtain \eqref{1stkkd}.

\section{Obtaining 1st law in ST gravity for both the frames in extended phase space} \label{AppenST}
The variation of the action \eqref{SJNEW} yields
\begin{align}
\d(\mg \ML)=\mg\Big[\ME_{ab}\d g^{ab}+\ME_{(\phi)}\d\phi-\frac{1}{16\pi}\frac{\p\V(\phi,\Lambda)}{\p\Lambda}\d\L+\na_a\Psi^a(q,\d q)\Big]~. \label{VARSJ}
\end{align}
Here, $q=\{g_{ab},\phi\}$ and the expressions of $\ME_{ab}$, $\ME_{(\phi)}$ and $\Psi^a(q, \delta q)$ are provided as follows:
\begin{eqnarray}
&& \ME_{ab}=\frac{1}{16\pi}\Big[\phi G_{ab}+g_{ab}\na_i\na^i\phi-\na_a\na_b\phi-\frac{\o(\phi)}{\phi}(\na_a\phi)(\na_b\phi)+\frac{\o(\phi)}{2\phi}g_{ab}(\na^i\phi)(\na_i\phi)+\frac{1}{2} g_{ab}\V(\phi,\L)\Big]~.
\no 
\\
&& \ME_{(\phi)}=\frac{1}{16\pi} \Big[R-\frac{\o(\phi)}{\phi^2}(\na^i\phi)(\na_i\phi)+\frac{2\o(\phi)}{\phi}\square\phi-\frac{\p\V(\phi,\L)}{\p\phi}+\frac{1}{\phi}\frac{d\o(\phi)}{d\phi}(\na^i\phi)(\na_i\phi)\Big]~.
\no 
\\
&&\ \ \ \ \ \ \ \ \ \ \ \ \ \ \ \ \ \ \ \ \ \ \ \ \ \ \ \ \ \ \ \ \ \ \ \ \textrm{and}
\no 
\\
&& \Psi^a(q, \d q)=\frac{1}{16\pi}\Big[2\phi P^{ibad}(\na_b\d g_{id})-2P^{iabd}(\na_b\phi)\delta g_{id}-2\frac{\o(\phi)}{\phi}(\na^a\phi)\delta\phi-\frac{3}{2}g^{ij}(\na^a\phi)\d g_{ij}
\no 
\\
&&\ \ \ \ \ \ \ \ \ \ \ \ \ \ \ \ \ \ \ \ \ \ \ \ \ \ \ \ \ \ -3(\na_b\phi)\delta g^{ab}-3\na^a(\d\phi)\Big]~, \label{EXPLICITSJ}
\end{eqnarray}
where we have defined $P^{abcd}=(g^{ac}g^{bd}-g^{ad}g^{bc})/2$.
The general Bianchi-like identity, in this case, is provided as 
\begin{align}
\na_b \ME^{ab}=-\frac{1}{2}\ME_{(\phi)}(\na^a\phi)+\frac{1}{32\pi}\frac{\p\V(\phi,\L)}{\p\L}(\na^a\Lambda)~. \label{BIANCHISJ}
\end{align}
For diffeomorphism ($x^a\longrightarrow x^a+\xi^a$) invariance, the arbitrary variation becomes Lie variation in \eqref{VARSJ}. Then Substituting \eqref{BIANCHISJ} in \eqref{VARSJ}, one obtains the conserved offshell Noether current ($\J^a$), which is given as
\begin{align}
\J^a=\ML\xi^a+2\ME^{ab}\xi_b-\Psi^a(q, \lie q)~. \label{NCURRENTJOR}
\end{align}
This conserved Noether current can further be written as the divergence of two-ranked anti-symmetric potential, which is called Noether potential ($\J^a=\na_b\J^{ab}$), where the Noether potential is provided as\footnote{Note, that the expression of Noether current and potential is the same both for extended phase space approach and the non-extended one.} 
\begin{align}
\J^{ab}[\xi]=\frac{1}{16\pi}\Big[\na^a(\phi\xi^b)-\na^b(\phi\xi^a)\Big]~. \label{NOPJOR}
\end{align}
Using Wald's approach, as described in section \ref{SECECON}, one can obtain the first law as $\delta M=T\d S+\Omega_H\delta J+ V\d P$ with $T=\kappa/2\pi$ and
\begin{eqnarray}
&& \d S=\frac{\pi}{\kappa}\delta\int_{\mathcal{H}} d\Sigma_{ab}\J^{ab}~;
\no 
\\
&& \d M=\frac{1}{2}\int_{\partial c_{\infty}} [\d(d\Sigma_{ab}\J^{ab})-2d\Sigma_{ab}\xi^{[a}\Psi^{b]}(q, \d q)]\Big|_{\xi=\xi_{(t)}}~;
\no 
\\
&& \d J=-\frac{1}{2}\int_{\partial c_{\infty}} [\d(d\Sigma_{ab}\J^{ab})-2d\Sigma_{ab}\xi^{[a}\Psi^{b]}(q, \d q)]\Big|_{\xi=\xi_{(\phi)}}~.
 \label{SMJSJ}
\end{eqnarray}
The expression of pressure and volume (adopting the regularization prescription) in provided in section \ref{SECST}.

Let us now quickly move forward to the Einstein frame. The variation of the Lagrangian, as provided in \eqref{SE}, yields
\begin{align}
\d(\sqrt{-\tilde{g}}\t{\ML})=\sqrt{-\t{g}}\t{\ME}_{ab}\delta{\t{g}^{ab}}+\sqrt{-\t{g}}\t{\ME}_{(\t{\phi})}\delta{\t{\phi}}-\sqrt{-\t{g}}\frac{\p\U(\tilde{\phi},\bar\L)}{\p\bar\L}\d\bar\L+\sqrt{-\t{g}}\t{\na}_a\t{\Psi}^a (\t{q},\delta{\t{q}})~, \label{VAR2}
\end{align}
where $\t{q}\in \{\t{g}_{ab}, \t{\phi}\}$ and
\begin{align}
 \t{\ME}_{ab}=\frac{\tilde{G}_{ab}}{16\pi}-\frac{1}{2}\tilde{\nabla}_a\tilde{\phi}\tilde{\nabla}_b\tilde{\phi}+\frac{1}{4}\tilde{g}_{ab}\tilde{\nabla}^i\tilde{\phi}\tilde{\nabla}_i\tilde{\phi}+\frac{1}{2}\tilde{g}_{ab}\U(\tilde{\phi},\bar\L)~;
\no 
\\
\t{\ME}_{(\t{\phi})}=\t{\na}_a\t{\nabla}^a\tilde{\phi}-\frac{\p\U(\t\phi,\bar\L)}{\p\tilde{\phi}}~;\ \ \ \ \ \ \ \ \ \ \ \
\no 
\\
\textrm{and} \ \ \ \ \ \ \ \ \ \ \ \ \ \ \ \ \ \ \ \ \ \ \ \ \
\no 
\\
\t{\Psi}^a (\t{q},\delta{\t{q}})=\frac{\delta\tilde{v}^a}{16\pi}-(\tilde{\nabla}^a\tilde{\phi})\delta\tilde{\phi}~. \ \ \ \ \ \  \label{EXACTEXEIN}
\end{align}
where $\delta \t{v}^a=2\t{p}^{ibad}\t{\nabla}_b \delta \t{g}_{id}$ and $\t{p}^{iabd}= (\t{g}^{ib}\t{g}^{ad}-\t{g}^{id}\t{g}^{ab})/2$~. The, Bianchi-like identity is provided as 
\begin{align}
\t{\na}_b\t{E}^{ab}=-\frac{1}{2}(\t{\na}^a\t{\phi})\t{E}_{\phi}+\frac{1}{2}\frac{\p\U(\tilde{\phi},\bar\L)}{\p\bar\L}(\t\na^a\bar\L)~. \label{RELEEE}
\end{align}
With these relations, one can obtain the conserved offshell Noether current in Einstein frame due to the diffeomorphism invariance ($\t x^a\longrightarrow\t x^a+\t\xi^a$) as
\begin{align}
\t{J}^a=\t{\ML}\t{\xi}^a+2\t{\ME}^{ab}\t{\xi}_b-\t{\Psi}^a (\t{q},\lie \t{q})~. \label{JEINT}
\end{align}
and, the Noether potential is provided as
\begin{align}
\t{J}^{ab}=\frac{1}{16\pi}(\tilde{\nabla}^a\tilde{\xi}^b-\tilde{\nabla}^b\tilde{\xi}^a)~. \label{JABEIN}
\end{align}
Again, following Wald's formalism, we obtain the first law as $\delta \t M=\t T\d \t S+\t\Omega_H\delta \t J+ \t V\d \t P$ where $\t T=\t\kappa/2\pi$ and 
\begin{eqnarray}
&& \d \t{S}=\frac{\pi}{\t{\kappa}}\delta\int_{\mathcal{H}} d\t{\Sigma}_{ab}\t{J}^{ab}~;
\no 
\\
&& \d \t{M}=\frac{1}{2}\int_{\partial c_{\infty}}[\d( d\t{\Sigma}_{ab}\t{J}^{ab})-2d\t\Sigma_{ab}\t{\xi}^{[a}\t{\T}^{b]}(\t{q}, \d \t{q})]\Big|_{\t{\xi}=\t{\xi}_{(t)}}~;
\no 
\\
&& \d \t{J}=-\frac{1}{2}\int_{\partial c_{\infty}} [\d(d\t{\Sigma}_{ab}\t{J}^{ab})-2d\t\Sigma_{ab}\t{\xi}^{[a}\t{\T}^{b]}(\t{q}, \d \t{q})]\Big|_{\t{\xi}=\t{\xi}_{(\phi)}}~.
\no 
\\
 \label{SMJTIL}
\end{eqnarray}
In Einstein frame, the pressure and thermodynamic volume (with regularization prescription) is defined in section \ref{SECST}. 
\\
{\underline{\textit{Equivalence of the thermodynamic parameters:}} One can obtain the Noether potentials are conformally connected as
\begin{align}
\bar \J^{ab}=\frac{\J^{ab}}{\phi^2} \label{JABrel}
\end{align}
and 
\begin{align}
\t \Psi^{a}(\bar q, \d\bar q)=\frac{\Psi^{a}(q, \d q)}{\phi^2}~. \label{PSIrel}
\end{align}
Also, one can obtain the elemental surface area of the two surface are connected in the two frames as $d\t{\Sigma}_{ab}=\phi^2d\Sigma_{ab}$. With the help of the above two relations \eqref{JABrel} and \eqref{PSIrel}, one obtains the mass, charge and angular momentum are equivalent in Jordan and Einstein frame. The equivalence of the temperature and angular velocity can be proved applying the exactly same argument as presented in section \ref{SECECON}, which we do not repeat here. The equivalence of pressure and volume is discussed in the section \ref{SECST}.


\end{document}